\documentstyle[twocolumn,aps,epsfig]{revtex}

\newcommand{\bqa}{\begin{eqnarray}}
\newcommand{\eqa}{\end{eqnarray}}
\newcommand{\nn}{\nonumber \\}

\begin{document}
\draft 
\twocolumn[\hsize\textwidth\columnwidth\hsize\csname @twocolumnfalse\endcsname
\title{On the origin of the hump structure in the  in-plane optical conductivity 
	of high $T_C$ cuprates based on a SU(2) slave-boson theory}

\author{Jae-Hyeon Eom, Sung-Sik Lee, Ki-Seok Kim and Sung-Ho Suck Salk$^a$}
\address{Department of Physics, Pohang University of Science and Technology,\\
Pohang, Kyungbuk, Korea 790-784\\
$^a$ Korea Institute of Advanced Studies, Seoul 130-012, Korea\\}
\date{\today}

\maketitle

\begin{abstract}
An improved version of SU(2) slave-boson approach is applied to study
the in-plane optical conductivity of the two dimensional systems 
of high $T_C$ cuprates.
We investigate the role of fluctuations of both the phase and  amplitude of order parameters
on the (Drude) peak-dip-hump structure in the in-plane conductivity as a function of hole
doping concentration and temperature.
The mid-infrared(MIR) hump in the in-plane optical conductivity is shown to originate
from the antiferromagnetic spin fluctuations of short range(the amplitude fluctuations
of spin singlet pairing order parameters), which is consistent with our previous  U(1) study.
However the inclusion of both the phase and amplitude fluctuations  is shown to substantially
improve the  qualitative feature of the optical conductivity by showing substantially reduced
Drude peak widths for entire doping range.
Both the shift of the hump position to lower frequency and the  growth of the hump peak height with
increasing hole concentration is shown to be consistent with observations.
\end{abstract}
\vskip2pc]
\narrowtext

\newpage

\section{INTRODUCTION}

 High $T_C$ superconductors are believed to be the systems of  strongly correlated
electrons essentially in two space dimension. 
This strong correlation results in the peak-dip-hump structure of the optical conductivity $\sigma(\omega)$ 
and  linear frequency  dependence of the scattering rate $1/\tau(\omega)$\cite{ROMERO,ROTTER,UCHIDA,PUCHKOV,LIU,TU}.
Both the appearance of the  hump(which occurs  at $\omega \approx 1000 cm^{-1}$ in YBCO\cite{UCHIDA,PUCHKOV} 
and Bi2212\cite{TU}) and the linear frequency dependence of the scattering rate($1/\tau(\omega) \sim \omega$)  
indicate  strong  deviations from the Drude model prediction of conventional Fermi-liquid.
Various theories have been proposed to explain these non-Fermi-liquid(NFL) like behavior 
in the charge dynamics(optical conductivity) of the high $T_C$ cuprates.
Using the nearly antiferromagnetic Fermi-liquid theory, Stojkovi$\acute{c}$ and Pines\cite{STOJKOVIC} reported a
study of normal state optical conductivity for the optimally doped and overdoped regions,
but not for the underdoped regions.
They showed that the highly anisotropic scattering rate in different regions of the Brillouin zone
leads to an average relaxation rate of the marginal Fermi-liquid  form by showing $1 / \tau (\omega)  \sim \omega $.
Their computed optical conductivity agreed well with experimental data for the normal state of an
optimally doped sample. 
Using the spin-fermion model\cite{MBP,CM} and spin susceptibility parameters obtained from
inelastic neutron scattering(INS) and nuclear magnetic resonance(NMR) measurements, Munzar et al. calculated the
in-plane optical conductivity of optimally doped YBCO\cite{MUNZAR}. 
They examined the  peak-dip-hump structure only at optimal doping and showed a good agreement with observation.
From the computed self energy they showed that the hump is originated from  the hot spot  and the Drude peak
from the cold spot  in the Brillouin zone.
Haslinger et al. reported  optical conductivities $\sigma (\omega)$ of optimally doped cuprates
in the normal state by allowing coupling between the fermions and the bosonic spin fluctuations\cite{HASLINER}.
They found that the width of the peak in spectral function $A_{\bf k}(\omega)$ scales linearly
with $\omega$ in both hot and cold spots and $\sigma (\omega)$ is inversely linear 
in $\omega$ up to very high frequencies.
 Besides these spin fluctuation theories mentioned here, by using a phenomenological form of the
charge collective mode(CM) propagator and empirical parameter values Caprara et al.
showed that charge-ordering instability is responsible for the hump in the optical conductivity
of high $T_C$ cuprates\cite{CAPRARA} limited only to the overdoped region.
The various above theories are reported in the limited range of hole doping 
resorting to empirical parameters by fitting INS or angle resolved photoemission spectroscopy(ARPES) measurements.
Thus the applicability of these theories may be limited only to a certain range of hole doping and temperature.
It is, thus, highly desirable to use a theory which does not depend on any  empirical parameters 
for the entire  range of hole doping and  temperature encompassing both the pseudogap phase and the
superconducting phase. 

Earlier Lee and Salk\cite{SSLEE} proposed  a   U(1) and SU(2) slave-boson  theories and 
showed an arch-shaped $T_C$ line in agreement with observations. 
Their theory is different from other previous slave-boson theories\cite{KOTLIAR,FUKUYAMA,WEN,GIMM} 
in that the Heisenberg term in the t-J Hamiltonian contains the contribution of 
coupling between the spin and charge degrees of freedom(the spinon pair and holon pair orders).
Recently, using the same U(1) slave-boson theory, we\cite{LEKS} were able to explain the peak-dip-hump 
structure of the observed optical conductivity, by showing that the hump is 
caused by the presence of spinon pairing order formed from the hot spot in the Brillouin zone.
In this paper, by using the SU(2) theory we report a rigorous examination on 
the origin of peak-dip-hump structures over the entire range of temperature($0<T<T_C$, $T_C<T<T^*$ and $T>T^*$) 
and the entire(underdoped, optimally doped and overdoped) range of hole doping.
In addition we discuss how the inclusion of the order parameter phase fluctuations markedly improve
the predicted in-plane optical conductivity over the previous results\cite{LEKS}.

\section{THEORY}

\subsection{U(1) AND SU(2) SLAVE-BOSON THEORIES OF t-J HAMILTONIAN}
In order to bring forth the essential physical points  we briefly  review the U(1) and SU(2) slave-boson theory
and point out the importance of coupling between the charge and spin degrees 
of freedom in determining the structure of the optical conductivity\cite{SSLEE}. 
The U(1) formulation is readily done by rewriting the electron operator as a composite of 
spinon($f$) and holon($b$) operators, $c_{i \sigma} = f_{i \sigma} b^{\dagger}_i$
with the single occupancy constraint, 
$\sum_{\sigma} f^{\dagger}_{i \sigma}f_{i \sigma} +b^{\dagger}_i b_i   =1$.
Then we obtain the partition function, 
\bqa
Z = \int 
{\cal D}f 
{\cal D}b
{\cal D}\lambda
 e^{-\int_{0}^{\beta} d \tau {\cal L}},
\eqa
with
 ${\cal L}= \sum_{i}\left( \sum_{\sigma}f^{\dagger}_{i \sigma} \partial_{\tau} f_{i \sigma} 
+  b^{\dagger}_{i} \partial_{\tau} b_{i}   \right)+ H_{t-J}
+ i \sum_{i}\lambda_i ( \sum_{\sigma} f^{\dagger}_{i \sigma}f_{i \sigma}+b^{\dagger}_i b_i -1)$ 
where $\lambda_i $  is the Lagrange multiplier field to enforce the single 
occupancy constraint and $H_{t-J}$, the U(1) slave-boson representation of the t-J
Hamiltonian, 
\bqa
H_{t-J} & = & -t\sum_{<i,j>, \sigma}(
f_{i\sigma}^{\dagger}f_{j\sigma}b_{j}^{\dagger}b_{i} 
+ c.c.) \nonumber \\
&& -\frac{J}{2} \sum_{<i,j>} b_i b_j b_j^{\dagger}b_i^{\dagger}
(f_{i\downarrow}^{\dagger}f_{j\uparrow}^{\dagger}-f_{i\uparrow}^ {\dagger}
f_{j\downarrow}^{\dagger})(f_{j\uparrow}f_{i\downarrow}-f_{j\downarrow}
f_{i\uparrow}) \nn
&& - \mu\sum_{i,\sigma}  f_{i\sigma}^{\dagger}f_{i\sigma}.
\label{Hamiltonian_fb}
\eqa
It is noted that the above Heisenberg interaction term(the second term) 
reveals coupling between the charge(holon) and spin(spinon) degrees of freedom.
Such coupling is ignored in other proposed theories\cite{KOTLIAR,FUKUYAMA,WEN,GIMM}.

Hubbard-Stratonovich transformations for the  hopping, spinon pairing and holon pairing
orders leads to  the partition function, 
\bqa
Z = \int
{\cal D}f 
{\cal D}b
{\cal D}\chi
{\cal D}\Delta^f
{\cal D}\Delta^b
{\cal D}\lambda
e^{-\int_{0}^{\beta} d \tau {\cal L}_{eff}},
\label{free_energy_final}
\eqa
where
 ${\cal L}_{eff} = {\cal L}_0 + {\cal L}_f + {\cal L}_b $
is the effective Lagrangian with
\small{
\bqa
{\cal L}_0 =&& \frac{J(1- x)^2 }{2} \sum_{<i,j>} \Big\{ 
|\Delta^f_{ij}|^2 + 
\frac{1}{2} |\chi_{ij}|^2 + \frac{1}{4} \Big\} \nn
&&+ \frac{J}{2} \sum_{<i,j>}
|\Delta^f_{ij}|^2 (|\Delta^b_{ij}|^2 + x^2), 
\label{orderparameter_u1}
\eqa }{\normalsize} for the order parameter Lagrangian and
\small{
\bqa
{\cal L}_f  
=&& \sum_{i,\sigma} f^\dagger_{i \sigma}( \partial_{\tau}- \mu^f ) f_{i \sigma} \nn
&& -\frac{J(1- x)^2}{4}\sum_{<i,j>,\sigma}
\Big\{ \chi^*_{ij} f^\dagger_{i \sigma}f_{j \sigma} + c.c. \Big\} \nn
&&- \frac{J(1- x)^2}{2} \sum_{<i,j>}
\Big\{ {\Delta^f}^*_{ij} ( f_{i , \downarrow} f_{j , \uparrow}
- f_{i , \uparrow} f_{j , \downarrow}) 
+ c.c. \Big\}
\label{spinon_u1}
\eqa }{\normalsize}for the spinon sector and  
\small{
\bqa
 {\cal L}_b =&& 
\sum_{i} b^\dagger_{i} (\partial_{\tau} - \mu^b) b_{i}
 -t \sum_{<i,j>}
\Big\{
\chi^*_{ij} b^\dagger_{i} b_{j} + c.c.
\Big\} \nn
&&-\frac{J}{2} \sum_{<i,j>}
|\Delta^f_{ij}|^2 \Big\{
{\Delta^b_{ij}}^* b_i b_j + c.c.
\Big\}
\label{holon_u1}
\eqa }{\normalsize}for
 the holon sector.
Here  $\chi$, ${ \Delta }^f$ and ${ \Delta }^b$ are
the hopping, spinon pairing and holon pairing order parameters
respectively. $\mu^f (\mu^b)$ is the spinon(holon) chemical potential and $x$, the hole concentration.

The SU(2) theory introduces the holon doublet
{\bqa
h_i = g_i \pmatrix{b_{i} \cr
0} \,\,\,  
=
\pmatrix{b_{1i} \cr
b_{2i}} \,\,\,  ,
\label{holon_spinor}
\eqa}and
 the electron operator is then written,
\bqa
c_{i \sigma} = \frac{1}{\sqrt{2}} h^{\dagger}_{i} \Psi_{i \sigma},
\label{su2_electron}
\eqa
with the spinon doublet 
$\Psi_{  i 1} =
g_i \pmatrix{f_{ i \uparrow} \cr
f^{\dagger}_{i \downarrow }} \,\,\, =
\pmatrix{f_{i 1} \cr
f^{\dagger}_{i 2}} \,\,\,$ 
and 
$ \Psi_{  i 2} =
g_i \pmatrix{f_{ i \downarrow} \cr
- f^{\dagger}_{ i \uparrow }} \,\,\, =
\pmatrix{f_{i 2} \cr
-f^{\dagger}_{i 1}} \,\,\,$. 
Here $g_i$ is the SU(2) rotation matrix.
Inserting Eq.(\ref{su2_electron}) into  the t-J Hamiltonian
and following same procedure with U(1) case we obtain  the effective Lagrangian,
\bqa
{\cal L}^{SU(2)}_{eff} = {\cal L}_0 + {\cal L}_f + {\cal L}_b ,
\label{su2_lagrangian}
\eqa
where
\small{
\bqa
&&{\cal L}_0 =  \frac{J(1-x)^2}{2} \sum_{<i,j>} \left(
                |\Delta^f_{ij}|^2 + \frac{1}{2} |\chi_{ij}|^2 + \frac{1}{4} 
                \right) \nn
                &&~~~~~~+ \frac{J}{2} \sum_{<i,j>} 
                |\Delta^f_{ij}|^2 \left(
                \sum_{\alpha, \beta} |\Delta^b_{ij;\alpha \beta}|^2 + x^2
                \right),
\label{su2_holon_spinon_lagrangian_0}
\eqa 
\bqa
&&{\cal L}_f =  \sum_i \Psi^\dagger_{i } \partial_\tau \Psi_{i } \nn
                &&~- \frac{J(1- x)^2}{4} \sum_{<i,j>} \left(
                \Psi^\dagger_{i }
                \left(
                \begin{array}{cc}
                \chi^*_{ij} &  - 2\Delta^f_{ij} \\
                - 2{\Delta^f_{ij}}^* & - { \chi_{ij} }
                \end{array}
                \right)
                \Psi_{j  }+ c.c.
                \right), 
\label{su2_holon_spinon_lagrangian_f}
\eqa 
\bqa
&&{\cal L}_b = \sum_i h^\dagger_i (\partial_\tau - \mu )h_i  \nn
		&&~~~~~- \frac{t}{2} \sum_{<i,j>} \left(
		 h^\dagger_i \left(
		\begin{array}{cc}
		\chi^*_{ij} &  - \Delta^f_{ij} \\
		- {\Delta^f_{ij}}^* & - { \chi_{ij} }
		\end{array}
		\right)
		h_j	+ c.c.
		\right) \nn
		&&~~~~~ - \frac{J}{2} \sum_{<i,j>} \left(
		|\Delta^f_{ij}|^2 h^\dagger_i 
		\left(
		\begin{array}{cc}
		\Delta^b_{ij;11} & \Delta^b_{ij;12} \\
		\Delta^b_{ij;21} & \Delta^b_{ij;22} 	
		\end{array}
		\right) 
		\left(  h^\dagger_j  \right)^T + c.c.
		\right), 
\label{su2_holon_spinon_lagrangian_b}
\eqa}{\normalsize}with 
$\Psi_{  i } =
\pmatrix{f_{i 1} \cr
f^{\dagger}_{i 2}} \,\,\,$. 
It is quite encouraging to realize that the third term in Eqs.(\ref{holon_u1})
and (\ref{su2_holon_spinon_lagrangian_b}) manifests the composite nature of spinon pairing and 
holon pairing to allow for the formation of Cooper pairs.
For the comparison of U(1) and SU(2) theory,  we rewrite 
the holon hopping term(the second term in Eq.(\ref{su2_holon_spinon_lagrangian_b})) 
by using  Eq.(\ref{holon_spinor}) as
\bqa
&&h^\dagger_i \left(
                \begin{array}{cc}
                \chi^*_{ij} &  - \Delta^f_{ij} \\
                - {\Delta^f_{ij}}^* & - { \chi_{ij} }
                \end{array}
                \right)
                h_j   \nn
&&= \left(
\begin{array}{cc}
b^\dagger_{1 i} & b^\dagger_{2 i}
\end{array}
  \right)
 \left(
                \begin{array}{cc}
                \chi^*_{ij} &  - \Delta^f_{ij} \\
                - {\Delta^f_{ij}}^* & - { \chi_{ij} }
                \end{array}
                \right)
\left(
\begin{array}{c}
b_{1 j} \\
 b_{2 j}
\end{array}
  \right) \nn
&&=\left(
\begin{array}{cc}
b^\dagger_{i} & 0
\end{array}
  \right)
g^\dagger_i  g_i
 \left(
                \begin{array}{cc}
                {{\chi^{(1)}}_{ij}}^* &  - \Delta^{f(1)}_{ij} \\
                - {\Delta^{f(1)}_{ij}}^* & - { \chi^{(1)}_{ij} }
                \end{array}
                \right)
g^\dagger_j g_j \left(
\begin{array}{c}
b_{j} \\
 0
\end{array}
  \right) \nn
&&=  {{\chi^{(1)}}_{ij}}^*  b^\dagger_i  b_{j}, 
\label{su2_order_lagrangian_compair_u1_su2}
\eqa
where $\chi^{(1)}_{ij}$ and $\Delta^{f(1)}_{ij}$ are the U(1) mean field 
hopping and spinon pairing order parameters respectively.
Thus the SU(2) order parameter matrix is related to the U(1)  order parameter matrix 
$U^{U(1)}_{ij}= \left(
                \begin{array}{cc}
                {\chi^{(1)}_{ij}}^* &  - \Delta^{f(1)}_{ij} \\
                - {\Delta^{f(1)}_{ij}}^* & - { \chi^{(1)}_{ij} }
                \end{array}
                \right)$ 
by gauge transformation,
\bqa
U^{SU(2)}_{ij}= \left(
                \begin{array}{cc}
                \chi^*_{ij} &  - \Delta^f_{ij} \\
                - {\Delta^f_{ij}}^* & - { \chi_{ij} }
                \end{array}
                \right)
= g_i U^{U(1)}_{ij} g^\dagger_j.
\label{su2_order_orderparameter_compair_u1_su2}
\eqa
As a result, in the SU(2) theory the low energy order parameter fluctuations can be included and allows
 the appearance of holon doublet\cite{WEN}.
Such SU(2) treatment resulted in an arch shaped $T_C$ line with the prediction
of more realistic optimal doping rate than the U(1) theory\cite{SSLEE}.

\subsection{OPTICAL CONDUCTIVITY IN U(1) AND SU(2) SLAVE-BOSON THEORIES}

 We obtain the optical conductivity ${ \sigma} (\omega) $ of an isotropic 2-D medium
by evaluating the current response function ${ \Pi}_{xx} (\omega)$,
\small{
\bqa
{ \sigma} ( \omega ) 
= \left. \frac{ \partial { J}_x (\omega) }{ \partial { E}_x (\omega)} \right|_{ { E}_x =0} 
=- \frac{1}{ i \omega } 
\left. \frac{ \partial^2 F  }{ \partial { A_x}^2} 
\right|_{{ A_x} =0} 
= \frac{ { \Pi}_{xx} (\omega) }{ i \omega}.
\eqa }{\normalsize}Here 
${ J}_x$ is the induced current in the $x$ direction and
$E_x$, the external electric field.
$F$ is the free energy,  $A_x$, the electromagnetic field and 
$\omega$, the frequency of applied electric field.
The total response function,
${ \Pi}_{xx}={ \Pi_{xx}}^{P}+{ \Pi_{xx}}^{D}$ is the sum of the
paramagnetic response function given by the
current-current correlation function ${ \Pi_{xx}}^{P}({\bf r}^\prime -{\bf r}, t^\prime -t)
= <{ j_x}({\bf r}^\prime ,t^\prime)
{ j_x}({\bf r},t )>-<{ j_x}({\bf r}^\prime ,t^\prime)><{ j_x}({\bf r},t )>$ 
with the current operator ${ j_x}({\bf r},t) = it(c^\dagger_{{\bf r}+{\bf x}, \sigma}(t) c_{{\bf r},\sigma}(t )
- c^\dagger_{{\bf r},\sigma}(t) c_{{\bf r}+{\bf x}, \sigma}(t )   )$
and the diamagnetic response function
associated with the average kinetic energy,
${\Pi_{xx}}^{D}=<{ K}>=-t \sum_{{\bf r},\sigma}
(c^\dagger_{{\bf r}, \sigma} c_{{\bf r}+{\bf x}, \sigma} + H.C.)$\cite{DAGOTTO}.

Recently, using the U(1) slave-boson theory we\cite{LEKS} studied the current response function
contributed only from spinon singlet pair excitations(the amplitude fluctuations
of the spinon pairing order parameter) and U(1) gauge fields(phase of the hopping order parameter)
upto the second order, by using 
\small{
\bqa
\Pi &=& \Pi^b -   \sum_{\alpha , \beta =  a, |\Delta^f|}
                  {\Pi^b}_{A \alpha}  {( {\Pi^b }+ {\Pi^f} )}^{-1}_{\alpha \beta}   {\Pi^b}_{ \beta A}\nn
     &=&    \frac{\Pi^f \Pi^b}{\Pi^f + \Pi^b}
+ \frac{
\left( \Pi^b_{a \Delta} -
       \frac{\Pi^b_{a \Delta} + \Pi^f_{a \Delta}}
       {\Pi^b + \Pi^f} \Pi^b
\right)^2 }
{   2  \frac{   ( \Pi^b_{a \Delta } + \Pi^f_{a \Delta } )^2}
      { \Pi^b + \Pi^f }
- ( \Pi^0_{\Delta \Delta} + \Pi^b_{\Delta \Delta} + \Pi^f_{\Delta \Delta} )   },
\label{Pi}
\eqa}{\normalsize}where
$\Pi^{f}$($\Pi^{b}$) is the spinon(holon)
response function associated with the internal and external gauge field
(${\bf a}$ and ${\bf A}$);
$ \Pi^f_{a \Delta} \equiv - \frac{ \partial^2 F^{f} }
{ \partial a \partial |\Delta^f| }$($ \Pi^b_{a \Delta} \equiv - \frac{ \partial^2 F^{b} }
{ \partial a \partial |\Delta^f| }$)  is 
the spinon(holon) response function contributed from  both
the gauge fields and the spinon pairing field
 and $\Pi^f_{\Delta \Delta}$, $\Pi^b_{\Delta \Delta}$ and $\Pi^0_{\Delta \Delta}$,
the response function contributed from  the spinon pairing field.
The first term represents the Ioffe-Larkin rule\cite{IOFFE}   which results from
the back-flow condition associated with  gauge field fluctuations.
On the other hand the second term is attributed to the  spinon singlet pair excitations.
It is noted that both terms in Eq.(\ref{Pi}) contain  the effects of 
the coupling between the charge and spin degrees of freedom.

For completeness it is essential  to include the effects of 
the phase fluctuations of hopping and spinon pairing order parameters 
 in addition to the amplitude fluctuations of hopping 
and spinon pairing order parameters.
Thus we write  the  order parameter matrix,
\small{
\bqa
&&{ U}^{SU(2)}_{ij} = \left(
\begin{array}{cc}
{\chi^*}_{ij} & -{ \Delta^f }_{ij} \\
- {{\Delta^f}^* }_{ij}& - {{\chi } }_{ij}
\end{array}
\right)  \nn
&&= \sqrt{|{\chi}_{ij}|^2 + |{\Delta^f}_{ij}|^2}
\left(
\begin{array}{cc}
\frac{{\chi^*}_{ij}}{\sqrt{|{\chi}_{ij}|^2 + |{\Delta^f}_{ij}|^2}} 
& - \frac{{\Delta^f}_{ij} }{\sqrt{|{\chi}_{ij}|^2 + |{\Delta^f}_{ij}|^2}}\\
- \frac{ { {\Delta^f}^* }_{ij} }{\sqrt{|{\chi}_{ij}|^2 + |{\Delta^f}_{ij}|^2}} 
& - \frac{ { {\chi } }_{ij}}{\sqrt{|{\chi}_{ij}|^2 + |{\Delta^f}_{ij}|^2}}
\end{array}
\right)  \nn
&&= { \eta }_{ij}
\left(
\begin{array}{cc}
{  e^{- i { \alpha}_{ij}} \cos \theta }_{ij} & - e^{i { \beta}_{ij}} \sin { \theta }_{ij} \\
- e^{-i { \beta}_{ij}} \sin { \theta }_{ij} & - e^{i { \alpha}_{ij}} \cos { \theta }_{ij} 
\end{array}
\right),
\label{22orderparameter}
\eqa}{\normalsize}where
${ \eta }_{ij} = \sqrt{|{\chi}_{ij}|^2 + |{\Delta^f}_{ij}|^2}$ is the `amplitude' of the
order parameter matrix, $ {\alpha}_{ij}$(${ \beta}_{ij}$), the phase of the hopping(spinon
pairing) order parameter and ${ \theta }_{ij}$, the relative phase angle of the hopping
and spinon pairing order parameters.
It is noted that $\alpha$, $\beta$ and $\theta$ represent the order parameter
phase fluctuations in the sense that they are low energy fluctuations. 
$\eta$ represents the amplitude fluctuations of the order parameters.

Allowing both the phase and amplitude fluctuations of the order
parameters, we obtain  the current response function,
\bqa
\Pi &=& \Pi^b -   \sum_{a , b =  \alpha , \beta , \theta , \eta }
                  {\Pi^b}_{A a}  {( {\Pi^b }+ {\Pi^f} )}^{-1}_{a b}   {\Pi^b}_{ b A}\nn 
   &=& \Pi^b -   \sum_{a , b =   \theta , \eta }
                  {\Pi^b}_{A a}  {( {\Pi^b }+ {\Pi^f} )}^{-1}_{a b}   {\Pi^b}_{ b A}  ,
\label{Pi_SU2}
\eqa
where
$ \Pi^f_{X Y} \equiv - \frac{ \partial^2 F^{f} }
{ \partial X \partial Y } \left(  \Pi^b_{X Y} \equiv - \frac{ \partial^2 F^{b} }
{ \partial X \partial Y } \right) $
with $X,Y = \alpha , \beta , \theta $ and $ \eta$. 
 Here we used the fact that $\Pi^b_{A \alpha }  = \Pi^b_{A \beta }  =0$ in the second line.
Therefore there is no coupling between the external electromagnetic field ${\bf A}$ and 
the phase fluctuation modes $\alpha$ and $\beta$. 
For details, we refer readers to Appendix A.

\section{COMPUTED RESULTS OF IN-PLANE OPTICAL CONDUCTIVITY}
\begin{figure}
\centerline{\epsfig{file=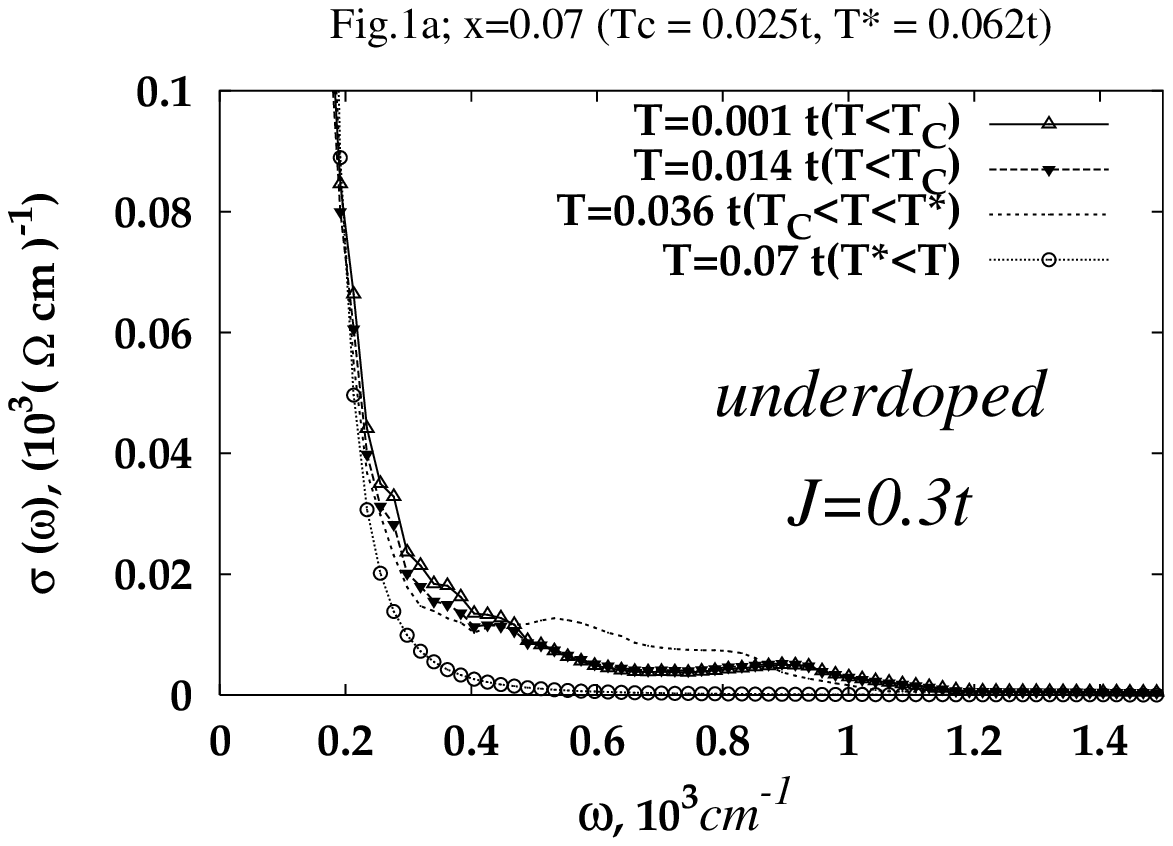,width=9cm,height=5cm, angle=0}}
\centerline{\epsfig{file=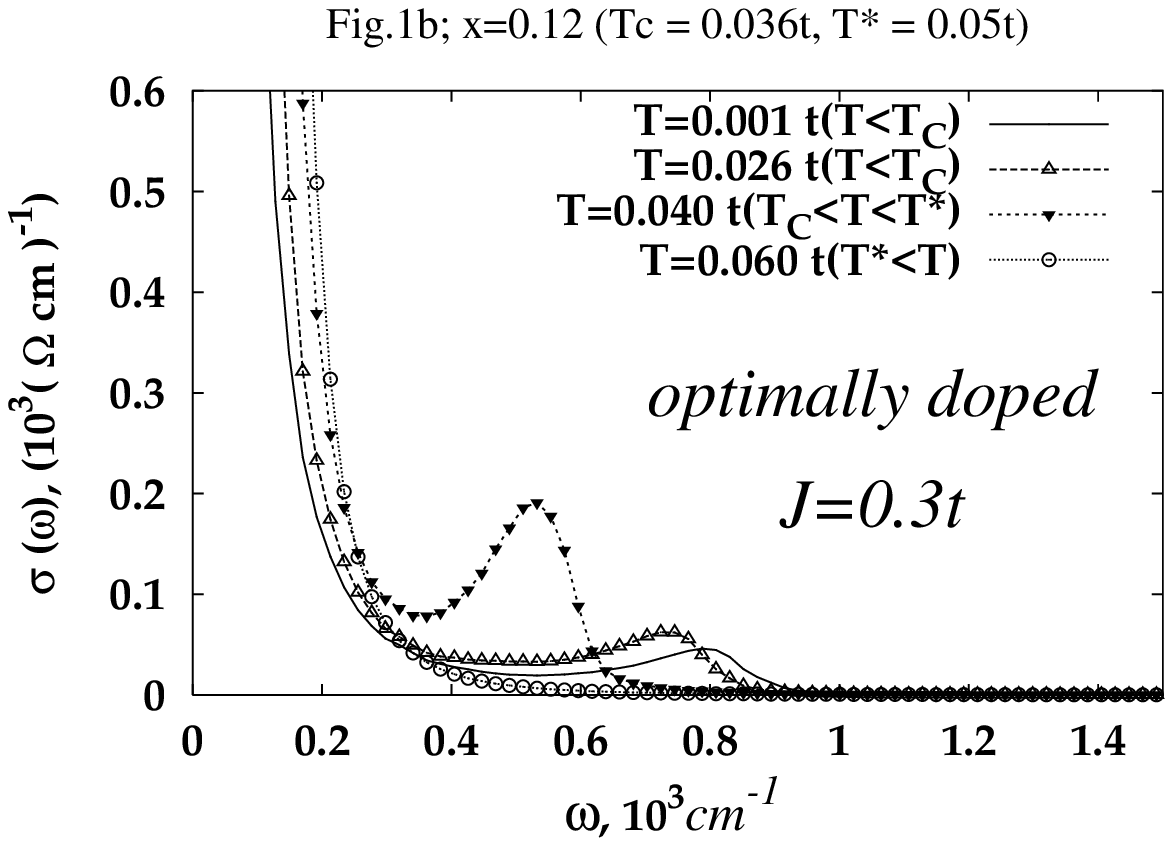,width=9cm,height=5cm, angle=0}}
\centerline{\epsfig{file=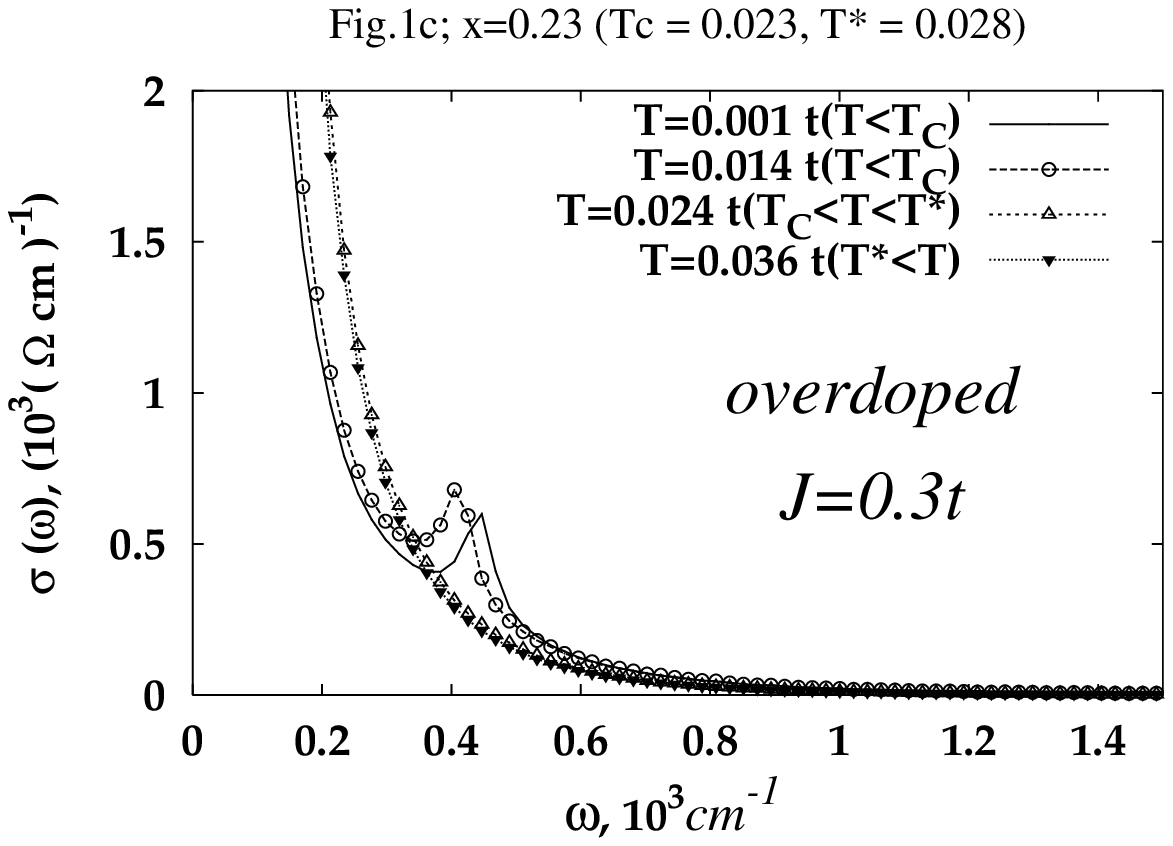,width=9cm,height=5cm, angle=0}}
\vspace{0.2cm}
\caption{Computed optical conductivities 
as a function of temperature 
for $x = 0.07$(underdoped), 
$x = 0.12$(optimally doped) and  $x = 0.23$(overdoped) cases 
with the antiferromagnetic Heisenberg coupling strength of J=0.3t\hspace{-0.2cm}}
\label{conductivity}
\end{figure}

We present the predicted optical conductivities
by allowing the order parameter fluctuations involving the parameters
 $\alpha$, $\beta$, $\theta$ and $\eta$ upto second order  in the framework of the SU(2) slave-boson theory 
of Lee and Salk\cite{SSLEE} by choosing J=0.3t for the underdoped($x=0.07$), optimally doped($x=0.12$)
and overdoped($x=0.23$) regions (Fig.\ref{conductivity}). 
Although not reported here, we find  no qualitatively marked changes with the variation of J
in the range of $0.1t<J<0.4t$.
Here  the bose condensation temperature($T_C$) and pseudogap temperature($T^*$) 
are given by our previous  SU(2) results\cite{SSLEE}. 
It is reminded that superconductivity is characterized by the order parameters of 
holon pair and spinon singlet pair. 
The spinon singlet pairs appear at temperature below $T^*$.
Below the superconducting temperature $T_C$ both the spinon singlet pair and holon pair order exist.
The predicted results show the  (Drude)Peak-dip-(MIR) hump structure 
at temperature below the pseudogap temperature $T^*$($T<T^*$) 
and  the hump disappears  above $T^*$($T^*<T$) 
indicating that the hump is originated from the spin singlet pair excitation, 
namely the antiferromagnetic spin fluctuations of the shortest 
possible correlation lengths.
Here we would like to point out that the predicted trend of both the shift of the hump position
to lower frequency and the growth of the hump peak height with increasing hole concentration
is consistent with observations\cite{MIHAILOVIC,ORENSTEIN}.
However, quantitative results in
hump position shift and hump peak height fail to agree with observations\cite{UCHIDA,PUCHKOV}.

\begin{figure}
\centerline{\epsfig{file=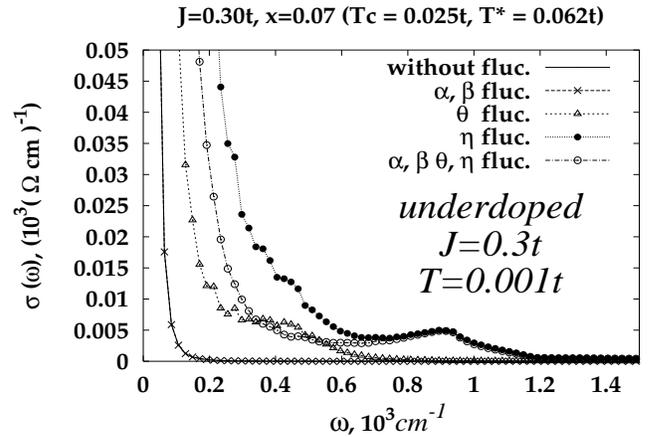, width=9cm, height=6cm,angle=0}}
\vspace{0.2cm}
\caption{Comparison of the optical conductivity 
1) without fluctuations,
with the fluctuations of 
2) $\alpha$,  $\beta$,  3) $\theta$, 
4) $\eta$ and  5) $\alpha$,  $\beta$, $\theta$, $\eta$; Result of case 1) and 2) is exactly same by virtue of 
$\Pi^b_{A \alpha }  = \Pi^b_{A \beta }  =0$(See Appendix A)}
\label{seperately}
\end{figure}

To thoroughly examine the origin of the 
peak-dip-hump structure, below we discuss predictions made by 
systematic changes of various phase fluctuation modes of $\alpha$ , $\beta$, and $\theta$.
Here $\alpha$ and  $\beta$  are the phases  of hopping and spinon 
pairing order parameters respectively and $\theta$, the relative phase angle between the two order 
parameters(see Eq.(\ref{22orderparameter})).
Fig.\ref{seperately} above is a replot of Fig.\ref{conductivity}a only for the case of $T=0.001t$.
Without the inclusion of order parameter fluctuations only Drude peak appears.
The same result is obtained with the allowance of only $\alpha$ and $\beta$ fluctuations. 
Thus the computed optical conductivities  with and without the phase fluctuations
of $\alpha$ and  $\beta$ are identical. This is    because 
$\Pi^b_{A \alpha }  = \Pi^b_{A \beta }  =0$ owing to the SU(2) symmetry(see Appendix A for proof)
as shown in Fig.\ref{seperately} above. 
The `$\theta$' mode  fluctuations resulted in a broader Drude peak
and some absorption at $200 cm^{-1} < \omega < 600 cm^{-1}$. 
The $\theta$ mode fluctuations represent an excitation mode 
associated with the amplitude fluctuations of both the 
hopping and spinon pairing order parameters  which are massive excitations. 
The hump structure  did not appear  with the $\theta$ mode fluctuations.
Upto now we discussed the predicted optical conductivity associated with  the phase
fluctuations involved with $\alpha$, $\beta$ and $\theta$. 
We now investigate the amplitude fluctuations of the order parameters involved with 
the `$\eta$' mode with $\eta = \sqrt{| \chi |^2 + |\Delta^f|^2}$.
As shown in Fig.\ref{seperately} only the $\eta$ mode fluctuations give a peak-dip-hump 
structure in the optical conductivity. However  the width of the 
Drude peak is affected by the inclusion of 
 the low energy excitations $\alpha$, $\beta$ and  $\theta$.
The phase fluctuations cause further reduction in the width
of Drude peak as compared to the case with the inclusion of only $\eta$ fluctuations.
Although not shown here, we observe  that the phase  fluctuations in the SU(2) 
cause a substantial reduction in the width of the Drude peak compared to the 
U(1) result but the position of the hump remains nearly identical.
The SU(2) result is in closer agreement with observations\cite{UCHIDA,PUCHKOV}
compared to the U(1) case.
On the other hand, the height of the hump is found to be lower in the SU(2) case as compared to the U(1) case.
The hump structure is largely originated from the amplitude fluctuations of the $\eta$ mode.
Both the  U(1) and SU(2) showed that the hump is originated from the antiferromagnetic
spin fluctuations of short range order,
although we cannot deduce separate contributions from the two different amplitude($| \chi |$ and $|\Delta^f|$)
fluctuations in this SU(2) calculation. 
This is because the amplitude fluctuations of both hopping order and spin(spinon) pairing order parameters 
occur in the background of antiferromagnetic fluctuations.

\begin{figure}
\centerline{\epsfig{file=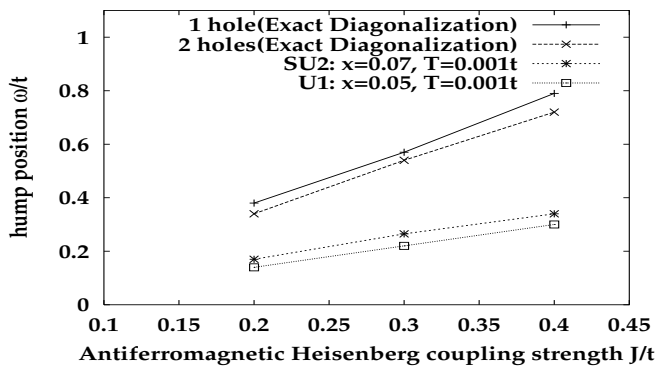, width=9cm, height=5cm, angle=0}}
\vspace{0.2cm}
\caption{Antiferromagnetic coupling dependence of 
hump position  for comparison
with the 
exact diagonalization 
calculations with two hole for $4 \times 4$ lattice at 
$T=0K$
and our slave-boson results 
at $x=0.07$(SU(2)) and $0.05$(U(1))
 at $T<<T_C$, that is $T=0.001t$}
\label{Hump_J}
\end{figure}

\begin{figure}
\centerline{\epsfig{file=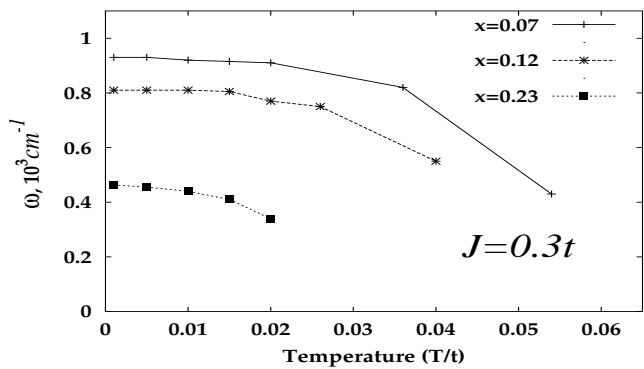, width=9cm, height=5cm,angle=0}}
\vspace{0.2cm}
\caption{Temperature dependence of 
hump position 
as a function of 
 hole concentration $x$ with antiferromagnetic coupling J=0.3t.}
\label{Hump}
\end{figure}

For an additional check on the origin of the  hump structure in a different aspect
 we computed the optical conductivity using  the Lanczos exact diagonalization method for a two hole doped
$4 \times 4$ lattice for various  values of Heisenberg antiferromagnetic coupling strength J.
The  peak-dip-hump structure appears only for non-zero values of J
and the position of the hump increases linearly with J.
This strongly indicates that  the hump is originated from the spin-spin correlations.
This trend is in agreement with our present calculations.
In Fig.\ref{Hump_J} we show the computed results of the hump position as a function of the
antiferromagnetic coupling strength J for comparison between
the Lanczos exact diagonalization method(with one and two hole doped cases 4$\times$4 square lattices) and 
U(1) and the SU(2) slave-boson theories. 
We note that the peak locations of the hump 
 obtained for both cases are sensitive to the variation of the antiferromagnetic 
coupling strength J, by showing a linear increase. 
This implies that the antiferromagnetic spin fluctuations are responsible for
the appearance of the hump structure.
Thus we conclude that the hump structure in both cases is originated from the
antiferromagnetic spin fluctuations, accompanying the amplitude fluctuations of the
spinon pairing order parameter at the  hot spots.
For comparison with observation we now present the predicted hump position
with the variation of hole concentration and temperature.
In Fig.\ref{Hump} the  peak position of hump is seen to remain nearly constant 
as a function of temperature far below $T^*$;
$T^*=0.062t$ for $x=0.07$, $T^*=0.05t$ for $x=0.12$ and $T^*=0.028t$ for $x=0.23$.
 In general, the predicted hump position
tends to shift to a lower frequency with increasing hole concentration
and with temperature, in agreement with the optical conductivity measurements\cite{UCHIDA,MIHAILOVIC,ORENSTEIN}. 
As shown in Fig.\ref{conductivity}, at temperature close to the pseudogap
temperature the hump structures
are seen to too rapidly disappear with increasing temperature compared to observations. 
This implies  that the antiferromagnetic spin fluctuations of short range order other than the spin
singlet pair(excitations) are important even  above $T^*$. 
Such antiferromagnetic fluctuations of short range are consistent
with NMR measurements\cite{YASUOKA} and other theoretical 
studies\cite{STOJKOVIC,MUNZAR,HASLINER}. 
\vspace{-0.2cm}

\section{SUMMARY}
In the present study, we studied  qualitative aspects of physics involved with the optical conductivity 
for the two-dimensional systems of strongly
correlated electrons for a wide range of both hole doping(under-,
optimal and over-doping) and temperature($T<T_C$, $T_C<T<T^*$,
and $T^*<T$) without resorting to any empirical parameters.
By observing the linear increase of the hump position
with the antiferromagnetic Heisenberg coupling strength J,
we showed that the hump structure in the optical conductivity
is originated from the antiferromagnetic
spin fluctuations which accompany the spin(spinon) singlet pair 
excitations(i.e., the antiferromagnetic fluctuations of the 
shortest possible correlation length) and  that the AF fluctuations 
of larger correlation lengths are also important for causing the hump structure.
This observation is consistent with NMR measurements\cite{YASUOKA} and other theoretical 
studies\cite{STOJKOVIC,MUNZAR,HASLINER}.
 The shift of the hump position to lower frequency and 
the increase of the hump peak height with increasing hole concentration
is see to be consistent with observations.
We have demonstrated that the $\theta$ mode phase fluctuations which represents the 
amplitude fluctuations of both  the hopping and spinon pairing order parameters
sharpens the Drude peak of the optical conductivity, 
 yielding markedly better agreement with observations compared to the U(1) theory  and that 
 the hump structure is not affected by the $\alpha$, $\beta$ and $\theta$ mode fluctuations, 
that is, the phase fluctuations involved  with the hopping and spin singlet pairing order parameters.
As pointed out earlier, our computed results fail to agree with observations 
in the magnitude of hump height. To achieve quantitative accuracy in the future study it is desirable to include the 
next nearest hopping term($t'$ term) in the t-J Hamiltonian and the 
effects of the AF fluctuations of larger correlation lengths.

\section{ACKNOWLEDGEMENT}
We would like to thank Young-Il Seo and Jae-Gon Um for helpful  discussions. 
One of us (SHSS) acknowledges the generous supports
of Korea Ministry of Education (Hakjin Excellence Leadership Program 2003)
and POSTECH Basic Science Institute  at Pohang Unversity of Science and Technology

\renewcommand{\theequation}{A\arabic{equation}}
\setcounter{equation}{0}
\section*{APPENDIX A: PROOF OF $  \Pi^b_{A \alpha }  = \Pi^b_{A \beta }  =0 $ IN THE SU(2) THEORY}
 From Eq.(\ref{su2_holon_spinon_lagrangian_b}), the holon action in the
presence of the external electromagnetic field is written,
\footnotesize{
\bqa
&&S_b (h,\alpha,\beta,\theta) = \int_0^\beta d \tau \sum_i h_i^{ \dagger}
			        ( \partial_\tau - \mu )  h_i \nn
&&			     	-\frac{t}{2} \sum_{<i,j>}
				\Big\{  
				e^{i A_{ij}}
				\eta_{ij}  
				 h^{ \dagger}_i
				\left(
				\begin{array}{cc}
				{  e^{-i { \alpha}_{ij}} \cos \theta }_{ij} & 
				- e^{i { \beta}_{ij}} \sin { \theta }_{ij} \\
				- e^{-i { \beta}_{ij}} \sin { \theta }_{ij} & 
				- e^{i { \alpha}_{ij}} \cos { \theta }_{ij}
				\end{array}
				\right)
				h_{j}
				+ c.c.
				\Big\} \nn
		             &&	-\frac{J}{2} \Delta^b  \sum_{<i,j>}
				\left\{   \eta^2_{ij} \sin^2 { \theta }_{ij}
				h_i^{ \dagger}
				\left(
                               \begin{array}{cc}
				1 & 0 \\
				0 & -1
                               \end{array}
				\right)
				\left(
				h_{j}^{ \dagger}
				\right)^T
				+ c.c.
				\right\}, 
\label{app_holon_action}
\eqa
}{\normalsize}where
we used $ {\chi}_{ij} = { \eta }_{ij} e^{i { \alpha}_{ij}} \cos \theta_{l}$   
and  ${\Delta^f}_{ij} = { \eta }_{ij} e^{i { \beta}_{ij}}  \sin \theta_{ij}$ . 
 Under the local SU(2) transformation holon field and the order parameters are transformed as
\footnotesize{
\bqa
&&h_i' =  {\bf g}_i  h_i ,\\
&& {U'}_{ij} = {\bf g}_i {U}_{ij} {\bf g}^{\dagger}_j ,
\label{app_transf_1}
\eqa \normalsize
where ${\bf g}_i=e^{i \frac{\vec{\tau}}{2} \cdot \vec{\Theta}_i}$ is the SU(2) transformation matrix  and
 $U_{ij} = \left(
\begin{array}{cc}
{\chi}^*_{ij} & -{ \Delta^f }_{ij} \\
- {{\Delta^f}^* }_{ij}& - {{\chi } }_{ij}
\end{array}
\right)$, the order parameter matrix. 
Here $\vec{\tau}$ are Pauli spin matrices and $\vec{\Theta}_i$, the rotation angle.   
 Considering the SU(2) rotation of $\pm \pi$ around the 2-axis the SU(2) transformation matrix is given by 
${\bf g}_i = (-1)^{i_x + i_y}\left(
 \begin{array}{cc}
      0 & 1 \\
     -1 & 0
 \end{array}
\right)$ 
and $U_{ij}$  is transformed as
\scriptsize{
\bqa
&&{U'}_{ij}= 
{ \eta '}_{ij}
\left(
\begin{array}{cc}
  e^{-i { \alpha }_{ij}'} \cos {\theta }_{ij}' & - e^{i { \beta }_{ij}'} \sin { \theta }_{ij}' \\
- e^{-i { \beta }_{ij}'} \sin { \theta  }_{ij}' & - e^{i { \alpha }_{ij}'} \cos { \theta }_{ij} '
\end{array}
\right) \nn
&&=
{\bf g}^\dagger_i {U}_{ij} {\bf g}_j  \nn
&&=   
(-1)^{{i}_x + {i}_y}
\left(
\begin{array}{cc}
0 & 1 \\
-1 & 0
\end{array}
\right) 
{ \eta }_{ij} 
\left(  
\begin{array}{cc}
{  e^{-i { \alpha}_{l}} \cos \theta }_{l} &                                   
- e^{i { \beta}_{l}} \sin { \theta }_{l} \\
- e^{-i { \beta}_{l}} \sin { \theta }_{l} &                                    
- e^{i { \alpha}_{l}} \cos { \theta }_{l}
\end{array}
\right) \nn
&& \times
(-1)^{{j}_x + j_y}
\left(
\begin{array}{cc}
0 & -1 \\
1 & 0
\end{array}
\right) \nn
&&=
{ \eta }_{ij}
\left(
\begin{array}{cc}
{  e^{i {  \alpha}_{l}} \cos \theta }_{l} &                                   
- e^{-i { \beta}_{l}} \sin { \theta }_{l} \\
- e^{i { \beta}_{l}} \sin { \theta }_{l} &                                    
- e^{-i { \alpha}_{l}} \cos { \theta }_{l}
\end{array}
\right),
\label{app_transf_2}
\eqa
}\normalsize
where  $j= i+x$ or $j = i+y$ and
$(-1)^{{i}_x + i_y} \times (-1)^{{j}_x + j_y} = -1$.
Thus we obtain $\alpha_{ij}' =  - \alpha_{ij}$,
$\beta_{ij}' =  - \beta_{ij}$, $\theta_{ij}' =   \theta_{ij}$
and $\eta_{ij}' =   \eta_{ij}$ . 
Then the holon action is given by
\scriptsize{
\bqa
&&S_b (h, A, \alpha,\beta,\theta , \eta)  \nn
&& = \int_0^\beta d \tau \sum_i   h_i^{ \dagger}  {\bf g}^\dagger_i {\bf g}_i
                                ( \partial_\tau - \mu )  {\bf g}^\dagger_i {\bf g}_i h_i \nn
&&                              -\frac{t}{2} \sum_{<i,j>}
                                \hspace{-0.2cm } \Big\{  
                                e^{i A_{ij}}
                                \eta_{ij}  
                                 h^{ \dagger}_i {\bf g}^\dagger_i {\bf g}_i
                                \left( \hspace{-0.2cm }
                                \begin{array}{cc}
                                {  e^{-i { \alpha}_{ij}} \cos \theta }_{ij} & 
                                \hspace{-0.2cm } - e^{i { \beta}_{ij}} \sin { \theta }_{ij} \\
                                - e^{-i { \beta}_{ij}} \sin { \theta }_{ij} & 
                                \hspace{-0.2cm } - e^{i { \alpha}_{ij}} \cos { \theta }_{ij}
                                \end{array}  \hspace{-0.1cm }
                                \right)
                                {\bf g}^\dagger_j{\bf g}_j h_{j}
                                + c.c.
                                \Big\} \nn
                             && -\frac{J}{2} \Delta^b  \sum_{<i,j>}
                                \left\{   \eta^2_{ij} \sin^2 { \theta }_{ij}
                                h_i^{ \dagger} {\bf g}^\dagger_i {\bf g}_i
                                \left(
                               \begin{array}{cc}
                                1 & 0 \\
                                0 & -1
                               \end{array}
                                \right)
                                \left(
                                h_{j}^{ \dagger} {\bf g}^\dagger_j {\bf g}_j
                                \right)^T
                                + c.c.
                                \right\} \nn
&&= \int_0^\beta d \tau \sum_i {h'}_i^{ \dagger}
                                ( \partial_\tau - \mu )  {h'}_i \nn
&&                              -\frac{t}{2} \sum_{<i,j>}
                                \Big\{  
                                e^{i A_{ij}}
                                {\eta '}_{ij}  
                                 {h '}^{ \dagger}_i
                                \left( \hspace{-0.2cm }
                                \begin{array}{cc}
                                {  e^{-i { \alpha }_{ij}'} \cos \theta  }_{ij}' & 
                                \hspace{-0.2cm } - e^{i { \beta }_{ij}'} \sin { \theta }_{ij}' \\
                                - e^{-i { \beta }_{ij}'} \sin { \theta }_{ij}' & 
                                \hspace{-0.2cm } - e^{i { \alpha }_{ij}'} \cos { \theta }_{ij}'
                                \end{array} \hspace{-0.1cm }
                                \right)
                                {h '}_{j}
                                + c.c.
                                \Big\} \nn
                             && -\frac{J}{2} \Delta^b  \sum_{<i,j>}
                                \left\{   {\eta '}^2_{ij} \sin^2 { \theta '}_{ij}
                                { h '}_i^{ \dagger}
                                \left(
                               \begin{array}{cc}
                                1 & 0 \\
                                0 & -1
                               \end{array}
                                \right)
                                \left(
                                {h '}_{j}^{ \dagger}
                                \right)^T
                                + c.c.
                                \right\} \nn
&& = S^b (h', A, - \alpha ,- \beta ,\theta  , \eta ).
\label{app_holon_action_final}
\eqa
}
\normalsize
Thus we find that the holon free energy is the even function of $\alpha$  and $\beta$:
\bqa
F^b (A, \alpha,\beta,\theta , \eta) 
&=& - \frac{1}{ \beta } ln \int {\cal D}h  e^{- S^b (h,A, \alpha,\beta,\theta, \eta)} \nn
&=& - \frac{1}{ \beta } ln \int {\cal D}h'  e^{- S^b (h' , A , -\alpha,-\beta,\theta, \eta)} \nn
&=& F^b ( A, -\alpha,-\beta,\theta, \eta), 
\label{app_holon_free_energy}
\eqa
 where we used the fact that the holon free energy is invariant under the local SU(2) transformation.
It is then obvious that the holon current response function
$ \frac{ \partial^2 F^b  }{ \partial A \partial \alpha }$ 
satisfies
\bqa
\Pi^b_{A \alpha }
&=& - \left. \frac{ \partial^2 F^b (A,\alpha,\beta,\theta,\eta) }{ \partial A \partial \alpha } \right|_{\alpha=\beta=0} \nn
&=&  - \left.  \frac{ \partial^2 F^b (A,-\alpha,-\beta,\theta,\eta) }{ \partial A \partial  \alpha } \right|_{\alpha=\beta=0} \nn
&=& - \Pi^b_{A \alpha }.
\label{app_zero}
\eqa
Thus we obtain $ \Pi^b_{A \alpha }  =0$.
Similarly $  \Pi^b_{A \beta }  =0$.
 The physical meaning of $\Pi^b_{A \alpha } = \Pi^b_{A \beta }  =0$
is that there is no coupling between the external electromagnetic field ${\bf A}$ 
and the order parameter phase fluctuations $\alpha$ and $\beta$.
Thus $\alpha$ and $\beta$ mode fluctuations are not affected by 
the external electromagnetic field 
and  the external electromagnetic field can not screen the fluctuations of these modes.

\end{document}